\documentclass[aps,prd,twoside,twocolumn,superscriptaddress,floatfix,nofootinbib]{revtex4}
\usepackage{amsmath}
\usepackage{bm}
\usepackage{xcolor}

\newcommand\beq{\begin{equation}}
\newcommand\eeq{\end{equation}}
\newcommand\beqn{\begin{eqnarray}}
\newcommand\eeqn{\end{eqnarray}}

\newcommand{\ba}{\begin{eqnarray}}
\newcommand{\ea}{\end{eqnarray}}
\newcommand{\be}{\begin{equation}}
\newcommand{\ee}{\end{equation}}

\newcommand\lsim{\mathrel{\rlap{\lower4pt\hbox{\hskip1pt$\sim$}}
        \raise1pt\hbox{$<$}}}
\newcommand\gsim{\mathrel{\rlap{\lower4pt\hbox{\hskip1pt$\sim$}}
        \raise1pt\hbox{$>$}}}

\newcommand{\jcap}{{J.~Cosm.~Astrop.~Phys.}}

\newcommand{\aap}{{Astron.~Astrophys.}}
\newcommand{\apjl}{{Astrophys.~J.~Lett.}}
\newcommand{\apjs}{{Astrophys.~J.~Supp.}}

\newcommand{\mnras}{{Mon.~Not.~R.~Astron.~Soc.}}
\usepackage{graphicx}
\usepackage[large]{subfigure}
\usepackage{amssymb, amsmath}
\usepackage[amssymb]{SIunits}
\usepackage{epstopdf}
\usepackage{aas_macros}
\usepackage{natbib}

\defcitealias{Hill2014ACTPDF}{H14}
\defcitealias{Battaglia2012}{B12}

\begin{document}

\title{An Accurate Analytic Model for the Thermal Sunyaev-Zel'dovich One-Point PDF}
%%%%%%%%%%%%%%%%%%%%%%%%%%%%%%%%%%%%%%%%%%%%%%%%%%%%%%%%%%%%%%%%%%%%%%%%%%%
\author{Leander~Thiele\footnote{lthiele@perimeterinstitute.ca}}
\affiliation{Perimeter Institute for Theoretical Physics, Waterloo ON N2L 2Y5, Canada}
\author{J.~Colin Hill}
\affiliation{School of Natural Sciences, Institute for Advanced Study, Princeton, NJ, USA 08540}
\affiliation{Center for Computational Astrophysics, Flatiron Institute, New York, NY, USA 10003}
 \author{Kendrick~M.~Smith}
 \affiliation{Perimeter Institute for Theoretical Physics, Waterloo ON N2L 2Y5, Canada}
%%%%%%%%%%%%%%%%%%%%%%%%%%%%%%%%%%%%%%%%%%%%%%%%%%%%%%%%%%%%%%%%%%%%%%%%%%%
\begin{abstract}
Non-Gaussian statistics of late-time cosmological fields contain information beyond that captured in the power spectrum.  Here we focus on one such example: the one-point probability distribution function (PDF) of the thermal Sunyaev-Zel'dovich (tSZ) signal in maps of the cosmic microwave background (CMB).  It has been argued that the one-point PDF is a near-optimal statistic for cosmological constraints from the tSZ signal, as most of the constraining power in tSZ $N$-point functions is contained in their amplitudes (rather than their shapes), which probe differently-weighted integrals over the halo mass function.  In this paper, we develop a new analytic halo model for the tSZ PDF, discarding simplifying assumptions made in earlier versions of this approach.  In particular, we account for effects due to overlaps of the tSZ profiles of different halos, as well as effects due to the clustering of halos.  We verify the accuracy of our analytic model via comparison to numerical simulations.  We demonstrate that this more accurate model is necessary for the analysis of the tSZ PDF in upcoming CMB experiments.  The novel formalism developed here may be useful in modeling the one-point PDF of other cosmological observables, such as the weak lensing convergence field.
\end{abstract}
%\pacs{98.80.-k, 98.70.Vc}
\maketitle

%%%%%%%%%%%%%%%%%%%%%%%%%%%%%%%%%%%%%%%%%%%%%%
\section{Introduction}
\label{sec:intro}

Cosmological inference has traditionally focused on measurements of the power spectrum (or its real-space analogue, the two-point correlation function).  For a Gaussian random field, this approach is sensible, as the power spectrum contains all statistical information in the data.  The primary cosmic microwave background (CMB) temperature and polarization anisotropies are canonical examples of such Gaussian fields~\cite{Planck2018params,Planck2015NG,WMAP9params}.  However, although the initial conditions for cosmic structure formation captured in the CMB are consistent with Gaussianity, non-Gaussian features inevitably develop in the late-time universe, due to non-linear gravitational evolution and complex baryonic physics on small scales.  Thus, the information content of late-time cosmological data sets, e.g., weak gravitational lensing maps or maps of the galaxy distribution, is not completely captured by the power spectrum.  For highly non-Gaussian fields, the amount of additional information in higher-order statistics can be significant.

In Ref.~\cite{Hill2014ACTPDF} (hereafter~\citetalias{Hill2014ACTPDF}), it was argued that the one-point probability distribution function (PDF) is a near-optimal non-Gaussian statistic for cosmological inference from maps of the thermal Sunyaev-Zel'dovich (tSZ) effect.  The tSZ effect is the up-scattering of CMB photons to higher energies due to Thomson scattering off hot, free electrons, which produces a unique distortion in the energy spectrum of the CMB~\cite{ZS1969,SZ1970}.  The tSZ effect probes the integrated pressure of free electrons along the line-of-sight (LOS); thus it is a biased tracer of free electrons, due to its dependence on the product of the electron number density and temperature.  In particular, because of this temperature dependence, the tSZ signal is predominantly sourced by electrons in galaxy groups and clusters, where electrons are virialized to high temperatures.  As the distribution of such objects in our Hubble volume is nearly Poissonian, maps of the tSZ effect are extremely non-Gaussian, dominated by individual rare, bright sources.  \citetalias{Hill2014ACTPDF} proposed the tSZ one-point PDF as an efficient statistic with which to extract the information in this non-Gaussianity.

Non-Gaussian properties of the tSZ signal have long been used for cosmological constraints, in the more familiar guise of inferring parameters via measurements of the halo mass function, in which candidate clusters are identified via the tSZ effect, their existence is confirmed and redshifts are estimated via multi-wavelength follow-up observations, and their masses are estimated via follow-up observations and/or scaling relations~(e.g.,~\cite{Hasselfield2013,Planck2015clusters,Bocquet2018}).  However, this approach utilizes only the brightest tSZ objects in the sky, e.g., those with signal-to-noise ($S/N$) greater than some threshold in the map.  An alternative strategy has thus been developed in the past two decades, in which ``indirect'' statistics of the tSZ signal are directly utilized as cosmological probes, such as the power spectrum~(e.g.,~\cite{Komatsu-Seljak2002,Hill-Pajer2013,McCarthy2014,Horowitz-Seljak2017,Sievers2013,George2015,Planck2013ymap,Planck2015ymap}), bispectrum/skewness~\cite{Wilson2012,Bhattacharya2013,Crawford2014}), and cross-power spectra with gravitational lensing maps~(e.g.,~\cite{HS2014,vanWaerbeke2014,Hojjatti2017}).  In such applications, no individual galaxy clusters are identified; instead, these statistics utilize information in objects below the $S/N$ threshold for individual detection, modeling their properties at the ensemble level.  In particular, it has been shown that tSZ statistics beyond the power spectrum contain a significant amount of cosmological information beyond that contained in cluster counts (for current $S/N$ thresholds) or the power spectrum~\cite{Wilson2012,Bhattacharya2013,Hill-Sherwin2013,Hill2014ACTPDF}.

A unifying feature of these statistical tSZ analyses is that their cosmological constraining power arises almost entirely from the one-halo term.  In other words, these statistics are just indirect methods of counting halos, weighted in different ways.\footnote{Thermal SZ -- gravitational lensing cross-correlations are an exception to this statement, but even these statistics have only moderate sensitivity to the two-halo term in current data~\cite{HS2014,Hojjatti2017}.}  For example, the tSZ power spectrum is dominated by the one-halo term at all $\ell \gtrsim 50$~\cite{Komatsu-Kitayama1999,Komatsu-Seljak2002,Hill-Pajer2013}.  Thus, rather than deriving cosmological constraints from spatial clustering information, these statistics do so via the halo mass function.  In particular, due to the tSZ signal's bias toward electrons in high-mass (i.e., high-temperature) halos, these statistics probe the exponential tail of the mass function, which makes them very sensitive probes of the amplitude of fluctuations, $\sigma_8$~(e.g.,~\cite{Komatsu-Seljak2002,Wilson2012,Hill-Sherwin2013,Bhattacharya2013}).  Importantly, this sensitivity to $\sigma_8$ is almost entirely encoded in the amplitude of these statistics, rather than their shape; due to the one-halo term's dominance (as mentioned above), the shape encodes information about intracluster medium (ICM) physics (and weak dependence on non-$\sigma_8$ cosmological parameters), while the amplitude is directly connected to integrals over the halo mass function, and hence $\sigma_8$.  This underlies the argument presented in~\citetalias{Hill2014ACTPDF} that the one-point PDF is an optimal statistic for cosmological inference from the tSZ signal.  The one-point PDF effectively captures the information in the amplitude of all $N$-point functions (or zero-lag moments), at the expense of information contained in the shape of the $N$-point functions (we anticipate that some shape information could be restored by considering the PDF on multiple smoothing scales, but such issues are not the focus of this paper).  While not optimal for constraining ICM parameters, the tSZ PDF does allow for the breaking of degeneracies between these parameters and $\sigma_8$, due to the different dependence of each moment on these parameters and $\sigma_8$ (this is a generalization of the argument for parameter degeneracy-breaking using the tSZ two- and three-point functions presented in Ref.~\cite{Hill-Sherwin2013}).

The improved cosmological constraining power of the tSZ PDF over other tSZ statistics was demonstrated in practice in~\citetalias{Hill2014ACTPDF}, which analyzed data from the Atacama Cosmology Telescope (ACT) at 148 GHz (using some 218 GHz data for foreground control as well).  Compared to an analysis of the same data using the tSZ skewness~\cite{Wilson2012}, the error bar on $\sigma_8$ was decreased by a factor of two in~\citetalias{Hill2014ACTPDF} (yielding $\sigma_8 = 0.793 \pm 0.018$), simply due to the improved cosmological sensitivity of the tSZ PDF over the skewness alone.  However, the $S/N$ of the measurement was not high enough to simultaneously constrain cosmological and ICM parameters (though the latter were marginalized over in obtaining the final $\sigma_8$ constraint).   The Planck Collaboration subsequently applied the PDF statistic to an analysis of their component-separated tSZ map, obtaining $\sigma_8 = 0.77 \pm 0.02$~\cite{Planck2015ymap}.

However, the theoretical modeling that was developed for the tSZ PDF in~\citetalias{Hill2014ACTPDF} made several simplifying assumptions, limiting its utility in upcoming measurements (its sufficiency for the analysis of the ACT data in~\citetalias{Hill2014ACTPDF} was verified explicitly using end-to-end simulations).  The analytic halo model of~\citetalias{Hill2014ACTPDF} assumed that clusters were Poisson-distributed on the sky and did not overlap, allowing the tSZ PDF to be computed via a simple integral over the mass function, given a model for the tSZ profile of each halo.  The primary goal of this paper is to remove these simplifying assumptions and generalize the analytic model from~\citetalias{Hill2014ACTPDF}, thereby allowing its use in analyses of the tSZ PDF from ongoing and upcoming CMB experiments (e.g., Advanced ACT~\cite{Henderson2016}, SPT-3G~\cite{Benson2014}, Simons Observatory~\cite{SO2018}, and CMB-S4~\cite{CMBS4ScienceBook}).  The assumptions are related to the distribution of halos sourcing the tSZ signal.  In~\citetalias{Hill2014ACTPDF}, it was assumed that these halos were sufficiently rare that they never overlapped on the sky.  For massive clusters, this assumption is valid, but as the tSZ signal of progressively lower mass halos is included in the PDF, this assumption breaks down.  For an experiment with relatively high noise levels (e.g., $\gtrsim 20 \, \mu{\rm K}$-arcmin with $\sim$arcmin-scale beams), the assumption is valid, since low-mass clusters are subsumed into the noise.  However, current and upcoming CMB experiments have noise levels well below this threshold, necessitating an improved model.

The other assumption from~\citetalias{Hill2014ACTPDF} that we will discard in this analysis is the neglect of halo clustering effects.  These effects are relevant due to the LOS projection inherent in the tSZ signal.  Note that the one-point PDF of 3D cosmic fields (i.e., the ``voxel'' PDF) does not receive any clustering contributions: as long as halo exclusion is enforced, only the one-halo term is necessary to compute the 3D PDF in the formalism used in this paper.  For projected 2D fields, however, this is not true.  Due to halo clustering, there is an excess probability for two halos to overlap along the LOS, compared to the Poisson expectation.  For the tSZ field, we will find (Fig.~\ref{fig:PDF_theory} below) that the clustering effect is relatively weak, but for future extensions of this formalism to the weak lensing convergence field, we expect that it will be significant.\footnote{See, e.g., Refs.~\cite{Liu2016,Patton2017,Liu-Madhavacheril2018} for simulation-based analyses of the weak lensing one-point PDF.  Note that another important difference between the tSZ and weak lensing cases, which will require further theoretical work, is the existence of negative-signal regions in the latter (cosmic voids), whereas the tSZ effect is strictly positive.}

The remainder of this paper is organized as follows.  In~\S\ref{sec:theory}, we review the halo model formalism of~\citetalias{Hill2014ACTPDF} and the associated assumptions, before proceeding to generalize the model in \S\ref{sec:new} and compare the results to those obtained in the simpler approach.  In~\S\ref{sec:sims}, we compare results from our analytic halo model to those obtained from numerical simulations, demonstrating its validity and accuracy.  In~\S\ref{sec:origin}, we use the analytic model to investigate the physical origin of the tSZ PDF signal.  \S\ref{sec:parameters} presents the cosmological and ICM parameter dependence of the tSZ PDF.  We then include noise and non-tSZ foregrounds in~\S\ref{sec:exp}, and demonstrate the sufficiency of our new model for the analysis of upcoming, high-precision CMB data sets.  We conclude in~\S\ref{sec:outlook}.

Our fiducial cosmology is flat $\Lambda$CDM with dimensionless Hubble constant $h = 0.7$, matter density $\Omega_{\rm M} = 0.25$, spectral index $n_s = 0.96$, $\sigma_8 = 0.8$, baryon density $\Omega_{\rm B} = 0.043$, sum of the neutrino masses $\Sigma m_{\nu} = 0$ eV, and CMB temperature $T_{\rm CMB} = 2.726\,{\rm K}$.

%%%%%%%%%%%%%%%%%%%%%%%%%%%%%%%%%%%%%%%%%%%%%%

\section{Background and Previous Model}
\label{sec:theory}

\citetalias{Hill2014ACTPDF} introduced a novel, simple analytic model for the tSZ one-point PDF.  Here we review this model and its assumptions, laying the groundwork for the more accurate model derived in the following section.

The tSZ signal is quantified by the Compton-$y$ parameter, which measures the integrated pressure of free electrons along the LOS:
\begin{equation}
\label{eq.tSZdef}
y(\mathbf{n}) = \frac{\sigma_{\rm T}}{m_e c^2} \int_{\rm LOS} dr \, n_e(r,\mathbf{n}) k_{\rm B} T_e(r,\mathbf{n}) \,,
\end{equation}
where $\sigma_{\rm T}$ is the Thomson cross-section, $m_e c^2$ is the electron rest-mass energy, $r$ is physical distance along the LOS, $n_e$ is the electron number density, $k_{\rm B}$ is Boltzmann's constant, and $T_e$ is the electron temperature.  

The tSZ effect produces a non-blackbody distortion in the energy spectrum of the CMB, which is negative (positive) at frequencies below (above) $\approx 218$ GHz.  Defining the dimensionless frequency $x \equiv h_{\rm Pl}\nu/(k_{\rm B} T_{\rm CMB})$, where $h_{\rm Pl}$ is Planck's constant and $\nu$ is the photon frequency, the tSZ spectral function is given by:
\begin{equation}
\label{eq.tSZspectralfunc}
g(\nu) = x \coth(x/2) - 4 \,.
\end{equation}
The tSZ-induced fluctuations in the CMB temperature field are then given by:
\begin{equation}
\label{eq.tSZtemp}
\frac{\Delta T_{\rm tSZ}(\mathbf{n},\nu)}{T_{\rm CMB}} = g(\nu) y(\mathbf{n}) \,.
\end{equation}
In this work, we neglect relativistic corrections to the tSZ effect~(e.g.,~\cite{Nozawa2006}), which are important for massive, hot clusters; these effects must be included in an actual data analysis (as they were in~\citetalias{Hill2014ACTPDF}).  Throughout the rest of the paper, our results will generally be given either in terms of Compton-$y$ or in terms of the tSZ field at a reference frequency of 148 GHz, where $g(148 \, {\rm GHz}) = -0.97881$.  We will also typically denote the tSZ-induced temperature fluctuation as simply $T \equiv \Delta T_{\rm tSZ}$.

We denote the (differential) tSZ one-point PDF as $P(y)$.  The binned version of the PDF used in much of this work is given by
\begin{equation}
p_i = \int_{y_i}^{y_{i+1}} dy\,P(y) \,.
\end{equation}
The concept underpinning the~\citetalias{Hill2014ACTPDF} model is to note that $p_i$ quantifies the fraction of sky subtended by Compton-$y$ values in the range $[y_i,y_{i+1}]$.  Thus, for a single spherically symmetric halo with an azimuthally symmetric projected $y$-profile $y(\theta)$, this would correspond to the area in the annulus between $\theta(y_i)$ and $\theta(y_{i+1})$, where $\theta(y_i)$ is the angular distance from the center of the halo to the radius where $y(\theta) = y_i$.  If one then makes the approximation that halos sourcing the tSZ signal are sufficiently rare that they never overlap on the sky, the final result for the full tSZ PDF is simply given by adding up such annular area contributions from all halos:
\begin{equation}
\label{eq:H14}
p_i = \int dz \, dM \frac{\chi^2}{H} \frac{dn}{dM} \pi \left( \theta^2(y_{i}) - \theta^2(y_{i+1}) \right) + \delta_i\,(1-F_{\rm clust}) \,,
\end{equation}
where $\chi(z)$ is the comoving distance to redshift $z$, $H(z)$ is the Hubble parameter, $dn(M,z)/dM$ is the halo mass function (i.e., the number of halos of mass $M$ at redshift $z$ per unit mass and comoving volume), and $\theta(y,M,z)$ is the inverse function of $y(\theta,M,z)$ (i.e., the Compton-$y$ profile of a halo of mass $M$ at redshift $z$), $F_{\rm clust}$ is the total sky area subtended by halos (assuming some radial cutoff), $\delta_i$ is unity if $y=0$ lies in the bin and zero otherwise, and redshift and mass dependences have been suppressed in the equation for compactness.

Eq.~(\ref{eq:H14}) makes two strong assumptions: (1) halos sourcing the tSZ signal are rare enough that their projected $y$-profiles never overlap; (2) the clustering of these halos (which would make overlaps more likely) can also be neglected.  These assumptions were valid for the analysis of ACT data in~\cite{Hill2014ACTPDF}, where the noise level was sufficiently high that the PDF could be modeled considering only halos for which these approximations are true (due to these halos being massive and hence rare).  The remainder of this paper is focused on discarding these assumptions, thus yielding a more accurate and general model for analysis of the tSZ PDF in ongoing and upcoming CMB experiments.  We will not focus on the modeling of non-tSZ foregrounds and noise (which were considered in detail in~\citetalias{Hill2014ACTPDF}, and must be in any future analysis as well), but rather only on the modeling of the physical tSZ PDF signal.

In this work, we adopt the same models for the physical quantities underlying the tSZ PDF as used in~\citetalias{Hill2014ACTPDF}.
We compute the halo mass function $dn/dM$ and halo bias $b(M,z)$ using the fitting functions of Ref.~\cite{Tinker2010}.
We compute electron thermal pressure profiles using the fitting function given by Ref.~\cite{Battaglia2012} (hereafter~\citetalias{Battaglia2012}), in order to facilitate direct comparison with their hydrodynamical simulations (\S\ref{sec:hydro_sims}).
For convenience, we give the relevant formulae here.
The thermal gas pressure at $r = r_{200\rm c}$ is given by
\begin{equation}
P_{200\rm c} = \frac{200GM_{200\rm c}\rho_{\rm c}(z)\Omega_{\rm B}}{2\Omega_{\rm M}r_{200\rm c}},
\end{equation}
with $\rho_{\rm c}(z)$ the critical density. Defining $x\equiv r/r_{200\rm c}$ and the core scale length $x_{\rm c}$, the pressure profile is parameterized as
\begin{equation}
\frac{P_{\rm th}(x)}{P_{200\rm c}} = \Pi_0\frac{(x/x_{\rm c})^{\gamma}}{(1+(x/x_{\rm c})^{\alpha})^{\beta}}.
\end{equation}
While $\alpha$ and $\gamma$ are held fixed, the remaining parameters are taken to have the following mass and redshift dependence:
\begin{align}
\Pi_0(\tilde M,z) 		&= P_0 \tilde M^{0.154} (1+z)^{-0.758},\\
x_{\rm c}(\tilde M, z) 	&= x_{\rm c,0} \tilde M^{-0.00865} (1+z)^{0.731},\\
\beta(\tilde M, z)		&= \beta_0 \tilde M^{0.0393} (1+z)^{0.415},
\end{align}
where $\tilde M \equiv M_{200\rm c}/10^{14}M_{\odot}$.
We take the fiducial values $\alpha^{\rm fid} = 1$, $\gamma^{\rm fid} = -0.3$, $P_0^{\rm fid} = 18.1$, $x_{\rm c,0}^{\rm fid} = 0.497$, and $\beta_0^{\rm fid} = 4.35$.
In~\S\ref{sec:parameters} we consider the effect of changes in $\alpha$, $P_0$, and $\beta_0$ on the one-point PDF.
All masses in the remainder of this work are given in terms of $M \equiv M_{200\rm m}$ unless otherwise stated.
When needed, we convert to $M_{200\rm c}$ using the NFW profile~\cite{NFW1997} and the concentration-mass relation of Ref.~\cite{Duffy2008},
making use of the Colossus package~\cite{Colossus}.
We note that this concentration model is calibrated in a mass range narrower than
our integration boundaries, but since the contributions from the low- and high-mass ends are generally small as shown
in \S\ref{sec:origin}, we do not expect this to be a significant source of error.
In our fiducial result we apply a radial cutoff to the $y$-profiles at $r_{\rm out} = 2\,r_{\rm vir}$,
where we define the virial radius with a redshift dependent overdensity approximated according to Ref.~\cite{BryanNorman1998}.
We choose integration boundaries such that $10^{11}\leq M\,[h^{-1}M_{\odot}] \leq 10^{16}$, and
$0.005 \leq z \leq 6$.
With these boundaries all integrals are converged.

\section{New Model}
\label{sec:new}

%\kms{I reorganized this section so that the contents of 
%Appendices~\ref{app:derivation_PDF_single_slice},~\ref{app:direct_convolution},~\ref{app:derivation_PDF_clustered}
%have been moved into the main text.  I left these appendices unchanged, but if you like the text in its current
%form, the appendices can be deleted now.  I also made scattered minor changes to the writing (see comments below).
%I don't feel strongly about either the organization or the details of the writing, so please feel free to
%iterate on this section, and change things freely!}

In this section, we derive the main result of this paper: an analytic model for the one-point PDF $P(y)$, making no simplifying assumptions about non-overlapping or non-clustering properties of halos.
We will do this in three steps.  First, we calculate the PDF for a narrow bin in mass $M$ and redshift $z$,
such that overlaps can be neglected (\S\ref{sec:pdf_single_bin}).  Second, we combine the contributions from
all $(M,z)$-values accounting for overlaps, but with large-scale clustering of halos neglected (\S\ref{sec:pdf_unclustered}).
Finally, we show how to include halo clustering (\S\ref{sec:pdf_clustered}).

It is convenient to work in conjugate space by introducing the Fourier transform (FT) of the one-point PDF:
\begin{equation}
\label{eq:def_tildeP}
\tilde P(\lambda) \equiv \int dy\,P(y)\,e^{i\lambda y}.
\end{equation}
Note that a calculation of $\tilde P(\lambda)$ is equivalent to a calculation of $P(y)$.\footnote{This approach bears similarities to the traditional $P(D)$ analysis used in radio point source studies~\cite{Scheuer1957,Condon1974}; however, our method accounts for tSZ sources' non-trivial profiles, and the overlaps (and clustering) associated with these.  We also note related work focused on the PDF of the tSZ power spectrum bandpowers~\cite{Zhang-Sheth2007}.}

\subsection{PDF for a narrow mass-redshift bin}
\label{sec:pdf_single_bin}

Let us consider now a narrow bin in halo mass and redshift of width $dM\,dz$ centered around mass $M$ and redshift $z$,
in which the number of halos is sufficiently small that halo overlaps can be neglected (by ``overlaps'', we mean overlaps of the projected $y$-profiles of these objects on the sky).  %\kms{Changed from ``\ldots in which there is at most a single halo.''}
% \commentleander{I have slightly changed the argument here, because I think it is a bit easier to
%	understand and probably familiar to most people from the standard derivation of the
%	Poisson distribution.}
We will calculate $\tilde P(\lambda)$ considering only halos in this bin.

We define the angular halo density in the narrow bin:
\begin{equation}
\frac{dn}{d\Omega} = \frac{\chi^2(z)}{H(z)}\,\frac{dn(M,z)}{dM}\,dM\,dz
\end{equation}
and let $y_0(M,z,\theta)$ denote the $y$-profile of a halo with mass $M$ and redshift $z$,
where $\theta$ is the angular distance from the halo center.
We assume that the profile has a finite radius $\theta_{\rm max}$, i.e.,~$y_0(M,z,\theta)=0$ for $\theta > \theta_{\rm max}$.
This is simply the projection of the radial cutoff defined in~\S\ref{sec:theory}.

We write $\tilde P(\lambda)$ as an expectation value:
\begin{equation}
\tilde P(\lambda) = \big\langle \exp (i\lambda y(\mathbf{n})) \big\rangle,
\end{equation}
where the sky location $\mathbf{n}$ is fixed, and the expectation value runs over random
halo placements.
%\kms{In the next few lines, I presented the calculation of the expectation value
% a little differently (the previous derivation is preserved in Appendix~\ref{app:derivation_PDF_single_slice}).
% Please feel free to keep whichever version you think is clearest!}
Since we are neglecting overlaps, we can trade this expectation value with an integral
over the angular profile:
\begin{equation}
\tilde P(\lambda) = \Big( 1 - \frac{dn}{d\Omega} \pi \theta_{\rm max}^2 \Big) + \frac{dn}{d\Omega} \int_0^{\theta_{\rm max}} d\theta \, 2\pi\theta \, e^{i\lambda y(\theta)},
\label{eq:pdf_narrow1}
\end{equation}
where the first term is the probability that no halo overlaps sky location ${\mathbf n}$,
and the second term integrates over overlap locations.  Introducing the auxiliary quantity
\begin{equation}
\label{eq:def_tildeY}
\tilde Y(M,z,\lambda) \equiv \int d\theta\,2\pi\theta\,\left( e^{i\lambda y_0(M,z,\theta)} - 1 \right),
\end{equation}
we rewrite Eq.~(\ref{eq:pdf_narrow1}) in the form:
\begin{equation}
\label{eq:narrow_bin_expanded}
\tilde P(\lambda) = 1 + \frac{\chi^2(z)}{H(z)} \, \frac{dn(M,z)}{dM} \, \tilde Y(M,z,\lambda) dM \, dz.
\end{equation}
For reasons that will be apparent in the next section, it will be convenient to write this as:
\begin{equation}
\label{eq:PDF_single_slice}
\tilde P(\lambda) = \exp\left( \frac{\chi^2(z)}{H(z)} \, \frac{dn(M,z)}{dM} \, \tilde Y(M,z,\lambda) dM \, dz \right)
\end{equation}
which is equivalent for a differential bin $(dM \, dz)$.

In Appendix~\ref{app:equivalence}, we show how to obtain
the formalism used in~\citetalias{Hill2014ACTPDF} from Eq.~(\ref{eq:narrow_bin_expanded}),
under the assumption that halo overlaps can be neglected.

\subsection{PDF neglecting halo clustering}
\label{sec:pdf_unclustered}

Now we calculate the one-point PDF from all masses and redshifts,
accounting for halo overlaps, but neglecting large-scale halo clustering.

Suppose we define a large number of narrow mass-redshift bins
$(M_1,z_1), \cdots, (M_N,z_N)$.  If halo clustering is neglected,
then the total $y$-signal is the sum of independent contributions
from each bin.  Therefore, the complete PDF from all $N$ mass-redshift 
bins is then obtained by convolution:
\begin{equation}
\label{eq:direct_convolution}
P(y) = \lim_{N\rightarrow\infty} P_{M_1,z_1}(y) \otimes \cdots \otimes P_{M_N,z_N}(y).
\end{equation}
Taking Fourier transforms, the convolution becomes multiplication, and simplifies as follows:
\begin{align}
\tilde P(\lambda) 
&= \lim_{N\rightarrow \infty} \prod_{i=1}^N \tilde P_{M_i,z_i}(\lambda)  \nonumber \\
%&= \lim_{N\rightarrow \infty} \prod_{i=1}^N \exp\left( \frac{\chi^2(z_i)}{H(z_i)} \, \frac{dn(M_i,z_i)}{dM_i} \, \tilde Y(M_i,z_i,\lambda) dM_i \, dz_i \right)  \nonumber \\
&= \prod_{i=1}^{\infty} \exp\left( \frac{\chi^2(z_i)}{H(z_i)} \, \frac{dn(M_i,z_i)}{dM_i} \, \tilde Y(M_i,z_i,\lambda) dM_i \, dz_i \right)  \nonumber \\
% &= \exp\left( \lim_{N\rightarrow \infty} \sum_{i=1}^N \frac{\chi^2(z_i)}{H(z_i)} \, \frac{dn(M_i,z_i)}{dM_i} \, \tilde Y(M_i,z_i,\lambda) dM_i \, dz_i \right)  \nonumber \\
&= \exp\left( \sum_{i=1}^{\infty} \frac{\chi^2(z_i)}{H(z_i)} \, \frac{dn(M_i,z_i)}{dM_i} \, \tilde Y(M_i,z_i,\lambda) dM_i \, dz_i \right)  \nonumber \\ 
&= \exp\left( \int \frac{\chi^2(z)}{H(z)} \, \frac{dn(M,z)}{dM} \, \tilde Y(M,z,\lambda) dM \, dz \right)
\label{eq:PDF_unclustered}
\end{align}
where we have used Eq.~(\ref{eq:PDF_single_slice}) in the second line.
% We show in Appendix~\ref{app:direct_convolution} that direct application of Eq.~\ref{eq:direct_convolution}
% yields results equivalent to our formalism.
%\kms{Do we want to keep Fig.~\ref{fig:convolution} in Appendix~\label{app:direct_convolution}, 
% showing equivalence between real-space convolution and Fourier-space multiplication?
% I don't feel strongly but slightly prefer dropping this figure, since we do show comparisions 
% with simplified simulations, which is a stronger test.}

\subsection{Including Halo Clustering}
\label{sec:pdf_clustered}

Finally, we include the effect of halo clustering.
In this subsection, we will denote the ``unclustered'' PDF
found in Eq.~(\ref{eq:PDF_unclustered}) by $\tilde P_{\rm u}(\lambda)$,
and denote the ``clustered'' PDF by $\tilde P_{\rm cl}(\lambda)$.

The derivation proceeds in two steps. 
First, we compute the one-point PDF $\tilde P_{\delta}(\lambda, \mathbf{n})$
in a fixed realization of the linear density field $\delta_{\rm lin}(\mathbf{n},z)$.
Second, we average over realizations of the Gaussian field $\delta_{\rm lin}$
to obtain $\tilde P_{\rm cl}(\lambda)$.
(Note that $\tilde P_{\delta}$ depends on $\mathbf{n}$, since translation invariance is
broken by a particular realization of $\delta_{\rm lin}$, but $\tilde P_{\rm cl}$ is a
translation-invariant PDF as usual.)

The quantity $\tilde P_{\delta}(\lambda, \mathbf{n})$ can be obtained
from Eq.~(\ref{eq:PDF_unclustered}) for the unclustered PDF, by biasing
the halo mass function with the halo bias $b(M,z)$, i.e.,
\begin{align}
\label{eq:tildeP_f}
\tilde P_{\delta}(\lambda,\mathbf{n}) &= \exp \int dM\,dz\,\Big\{\frac{\chi^2(z)}{H(z)}\,\tilde Y(M,z,\lambda) \nonumber \\
							   & \quad\times \frac{dn(M,z)}{dM}[1+b(M,z)\delta_{\rm lin}(\mathbf{n},z)]\Big\}.
\end{align}
We note that this expression is only meaningful as long as we can define a sufficiently large environment around $\mathbf{n}$
in which $\delta_{\rm lin}(\mathbf{n}+\mathbf{n}',z) \simeq \delta_{\rm lin}(\mathbf{n},z)$.
This assumption is justified because $\delta_{\rm lin}(\mathbf{n},z)$ varies slowly in comparison to the typical cluster radius.

The PDF including clustering is obtained by averaging over realizations of the linear
density field $\delta_{\rm lin}(\mathbf{n},z)$ in Eq.~(\ref{eq:tildeP_f}):
\begin{equation}
\tilde P_{\rm cl}(\lambda) = \langle \tilde P_{\delta}(\lambda, \mathbf{n}) \rangle_{\delta(\mathbf{n},z)}.
\end{equation}
Introducing
\begin{align}
A(\lambda, \mathbf{n})       &\equiv \int dz\,\delta_{\rm lin}(\mathbf{n},z)\alpha(z,\lambda),  \label{eq:A_def} \\
			\alpha(z,\lambda)&\equiv\int dM\,b(M,z)\frac{\chi^2(z)}{H(z)}\frac{dn(M,z)}{dM}\tilde Y(M,z,\lambda),  
\end{align}
we write Eq.~(\ref{eq:tildeP_f}) as:
\begin{equation}
\tilde P_{\delta}(\lambda,\mathbf{n}) = \tilde P_u(\lambda) e^{A(\lambda, \mathbf{n})}.
\end{equation}
Using the identity $\langle e^x \rangle = e^{\langle x^2 \rangle / 2}$ for a Gaussian random variable $x$,
we obtain:
\begin{equation}
\tilde P_{\rm cl}(\lambda) = \tilde P_{\rm u}(\lambda)\exp \frac{1}{2}\langle A^2(\lambda, \mathbf{n}) \rangle_{\delta}.
\end{equation}
The expectation value $\langle A^2 \rangle$ can be evaluated
using the Limber approximation and the LOS integral
representation in Eq.~(\ref{eq:A_def}).  The result is:
\begin{equation}
\langle A^2(\lambda, \mathbf{n}) \rangle_{\delta} = \int dz\,H(z)D^2(z)\alpha^2(z,\lambda)\times\int\frac{k\,dk}{2\pi}P_{\rm lin}(k).
\end{equation}
where $P_{\rm lin}(k)$ is the linear matter power spectrum at $z = 0$ and
$D(z)$ is the growth factor with $D(0) = 1$.
This gives our final expression for the one-point PDF:
\begin{align}
\tilde P_{\rm cl}(\lambda) &= \tilde P_{\rm u}(\lambda) \exp \bigg( \frac{1}{2}
   \int dz\, H(z) D(z)^2 \alpha(z,\lambda)^2 \nonumber \\
 & \hspace{2.5cm} \times
   \int \frac{k\,dk}{2\pi}P_{\rm lin}(k) 
\label{eq:PDF_clustered}
\bigg)
\end{align}
where $\tilde P_{\rm u}(\lambda)$ is the ``unclustered'' PDF in Eq.~(\ref{eq:PDF_unclustered}).
%\kms{Previously the definition of $\alpha$ was ``expanded'' in Eq.~(\ref{eq:PDF_clustered}), but I thought it
% looked a little nicer with explicit factors of $\alpha$.}

\subsection{Quantifying the Effects of Overlaps and Clustering}
\label{sec:quantifying_overlaps_and_clustering}

%\commentleander{The following paragraph can potentiallly go somewhere else.
%	These are definitions and technicalities.}

In the following, we denote the PDF integrated over a given temperature bin by $p_T$ and evaluate the temperature decrement
corresponding to a specific $y$-signal at $\nu = 148\,{\rm GHz}$.
Unless otherwise stated, we bin the PDF into temperature bins of width $1\,\mu{\rm K}$.
The linear power spectrum $P_{\rm lin}(k)$ and growth factor $D(z)$ are computed using \texttt{CAMB}~\cite{CAMB}.
We briefly mention two numerical properties of our analytic method.
If the integration boundaries (in mass and cluster radius) are chosen too narrow,
the Fourier transform $\tilde P(\lambda)$ does not vanish for $\lambda \rightarrow \infty$.
This gives rise to ringing in the PDF.
Furthermore, the PDF at high $|T|$ is only properly converged if the $y$-profiles are
evaluated on a very fine angular grid, which is due to their rapid variation for angles
close to the clusters' center.

\begin{figure}
\includegraphics[width = 0.5\textwidth]{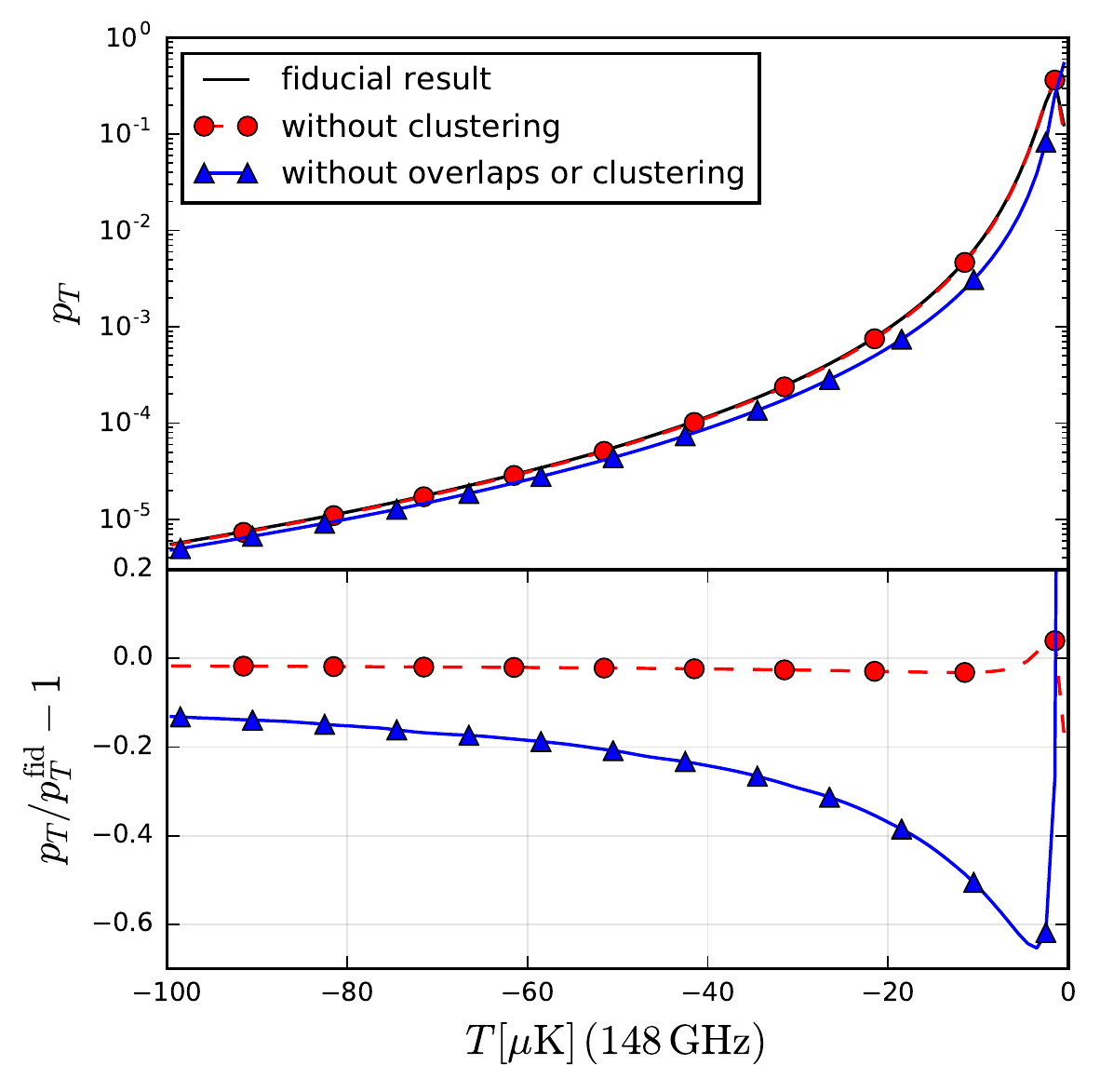}
\caption{\label{fig:PDF_theory} Effects of clustering and overlaps on the tSZ one-point PDF. The fiducial curve (black; including overlaps and clustering) and the result without clustering (red circles) are obtained using the new Fourier-based approach presented in \S\ref{sec:new}, while the result without clustering or overlaps (blue triangles) is calculated using the formalism of~\citetalias{Hill2014ACTPDF} as described in \S\ref{sec:theory}.  The bottom panel presents the fractional difference of the latter two curves with respect to the fiducial curve.  For clarity, only selected datapoints are plotted with markers.}
\end{figure}

We now turn to concrete results from our analytic model.
In Fig.~\ref{fig:PDF_theory} we show the effects of clustering and overlaps.
The fiducial result is obtained using Eq.~(\ref{eq:PDF_clustered}),
the result neglecting halo clustering is calculated from Eq.~(\ref{eq:PDF_unclustered}),
and the result neglecting both overlaps and clustering follows from the formalism
developed in~\citetalias{Hill2014ACTPDF} (Eq.~\ref{eq:H14}).
Overlaps have a much larger impact on the PDF, in particular for low $|T|$ values.
These temperature bins are dominated by numerous low-mass halos,
which have a larger probability to overlap.
Clustering increases the PDF for almost all $|T|$ values, but it is less important
than overlaps.
This is explained by the fact that it only plays a role if the gravitational interaction
aligns two nearby clusters along the line of sight, which has subdominant probability in comparison
to random alignments on arbitrarily large scales.
Clustering slightly decreases the PDF for $T > -4\,\mu\rm K$.
These bins are dominated by very numerous low mass halos,
which have a relatively high probability to gravitationally interact
and produce an alignment, which would push their contribution to higher $|T|$ values.
The lowest $|T|$ bin, on the other hand, sees an increase in the PDF due to clustering,
which is clear because clustering increases the clear sky fraction $(1-F_{\rm clust})$.  Note that the unphysical divergence of the~\citetalias{Hill2014ACTPDF} model in the lowest $|T|$ bin (arising from the neglect of overlaps) is removed by the new Fourier-based approach developed here.

%%%%%%%%%%%%%%%%%%%%%%%%%%%%%%%%%%%%%%%%%%%%%%
\section{Comparison to Simulations}
\label{sec:sims}

We check the validity of our analytic approach by comparing to the results of two different simulation methods.  First, we produce ``simplified''random  maps, in which unclustered halos are randomly distributed on a simulated flat-sky map and assigned Compton-$y$ profiles computed with the~\citetalias{Battaglia2012} pressure profile.  We then measure the average tSZ PDF from these maps.  By construction, the one-point PDF of our simplified simulations should agree perfectly with our ``unclustered'' analytic result in Eq.~(\ref{eq:PDF_unclustered}), but verifying the agreement is a strong check on details of the implementation, which are nontrivial (see~\S\ref{sec:quantifying_overlaps_and_clustering}).

Second, we measure the tSZ PDF from Compton-$y$ maps constructed directly from the cosmological hydrodynamics simulations of~\citetalias{Battaglia2012}.\footnote{We thank N.\ Battaglia for sharing these maps.}  The simulated maps in this case only include tSZ signal from halos at $z<1$, and thus in this section we set the upper redshift integration boundary to $z = 1$ in our analytic calculations, in order to enable direct comparison with the simulated maps.  Note that our fiducial cosmology is identical to that used in~\citetalias{Battaglia2012}.

\begin{figure}
\includegraphics[width = 0.5\textwidth]{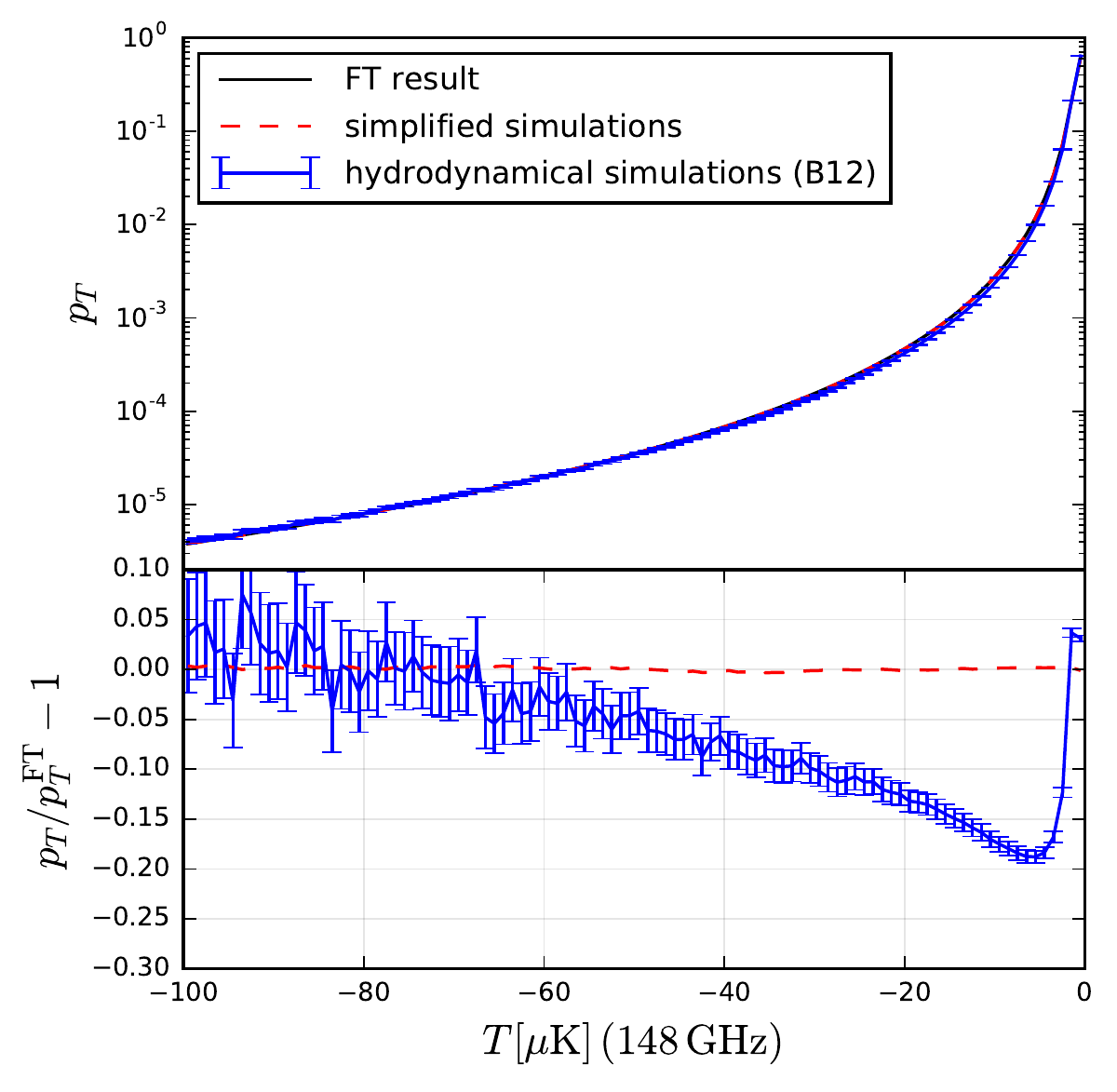}
\caption{\label{fig:PDF_sim} Comparison of our analytic tSZ PDF results to the PDF measured via two sets of simulations.  The top panel shows our fiducial analytic result computed via Eq.~\ref{eq:PDF_clustered} (black), the PDF measured from ``simplified'' simulations that neglect halo clustering (dashed red), and the PDF measured from cosmological hydrodynamics simulations (solid blue with error bars).  The bottom panel shows the fractional difference of the analytic results with respect to the simulation-derived results.  The discrepancy seen with the hydrodynamical simulations is investigated further in Fig.~\ref{fig:hmf_interpolations}, and found to be due to halo mass function differences.  Note that we compare the hydrodynamical simulations to the analytic model including clustering, while the simplified simulations are compared to the analytic model without clustering (not plotted in the top panel for clarity).  The error bars on the~\citetalias{Battaglia2012} results are estimated from the scatter amongst the simulated maps.  Note that in the upper panel the difference between the clustered and unclustered analytic results would be invisible by eye, and thus only one curve has been plotted.}
\end{figure}

\subsection{Simplified Simulations}
\label{sec:simplified_sims}

The simulations described in this section are produced as follows.
We construct individual maps of area $\Omega = 10\times10\,{\rm deg}^2$,
with square pixels of side-length $3\,{\rm arcsec}$.
We then consider discrete, narrow bins in mass and redshift of size
$dM\,dz$ centered around $M,z$, for which we compute the average number of such halos in the map via $n(M,z) = \Omega\,dn(M,z)/d\Omega$.
For each such mass-redshift bin, we populate the map with $y$-profiles (using the~\citetalias{Battaglia2012} pressure profile), whose number
is given by the probability distribution $w(\lceil\!n\!\rceil) = n - \lfloor\!n\!\rfloor$,
and $w(\lfloor\!n\!\rfloor) = \lceil\!n\!\rceil - n$. Since this distribution
reproduces the correct mean, it is valid to use it to find the
average tSZ PDF computed from many maps.
We find that sampling the number of halos from the physically more realistic Poisson distribution leads to relatively slow convergence
of the average PDF, but we have confirmed that it yields a consistent result with the more rapid approach.  Note that we do not include halo clustering in these maps.

Fig.~\ref{fig:PDF_sim} shows the average tSZ PDF computed from 507 simplified simulations (dashed red curve).  As the maps do not include clustering of halos, we compare the simulation-derived PDF to the analytic result from~\S\ref{sec:pdf_unclustered}, in which clustering effects are not included (but halo overlaps are).  The discrepancy with our analytic result is on average $\approx 0.2\%$ and decreases as more maps are added to the average.  This confirms the validity of our analytic formalism, in the limit where halo clustering effects can be neglected.

\subsection{Cosmological Hydrodynamics Simulations}
\label{sec:hydro_sims}

We now compare the results of the full analytic calculation presented in \S\ref{sec:pdf_clustered} (including halo clustering) to measurements of the tSZ PDF from cosmological hydrodynamics simulations.  We use Compton-$y$ maps constructed by direct LOS integration (to $z=1$) of randomly rotated and translated simulated volumes from Ref.~\cite{Battaglia2010}.  These are the same simulations from which the~\citetalias{Battaglia2012} pressure profile fitting function was extracted; thus, the comparison here is a direct test of our analytic formalism for the tSZ PDF, with no additional tuning of ICM parameters required.  The simulations were performed using the smoothed particle hydrodynamics code GADGET-2, with a custom implementation of a sub-grid prescription for feedback from active galactic nuclei.  The pressure profile model extracted from the simulations has subsequently been found to agree with a wide range of tSZ and X-ray measurements~(e.g.,~\cite{Sun2011,Planck2013stack,Hajian2013,HS2014,Greco2015,Hilletal2018}).  A full description of the simulations can be found in Ref.~\cite{Battaglia2010}.

We consider 390 Compton-$y$ maps extracted from the simulations, each of area $\Omega = 4.09\times4.09\,{\rm deg}^2$, with square pixels of side-length $6\,{\rm arcsec}$.  The average tSZ PDF measured from this suite of $y$-maps is shown in Fig.~\ref{fig:PDF_sim} (blue curve with error bars).  As mentioned above, an upper redshift cut at $z=1$ is applied in the construction of the maps, so as to minimize correlations arising from high-redshift objects common to multiple maps.  We apply this redshift cut in the analytic calculation in this subsection (and only this subsection) for consistency with the simulated maps.

The agreement in Fig.~\ref{fig:PDF_sim} between the tSZ PDF measured from the hydrodynamical simulations and our analytic result is much worse than the agreement for the simplified simulations seen in the previous subsection.  However, this difference can be traced back to the discrepancy between the halo mass function found in~\citetalias{Battaglia2012} and the mass function~\cite{Tinker2010} used in this work.  We explicitly confirm that this halo mass function difference can indeed give rise to discrepancies in the one-point PDF as observed in Fig.~\ref{fig:PDF_sim}.

\begin{figure}
\includegraphics[width = 0.5\textwidth]{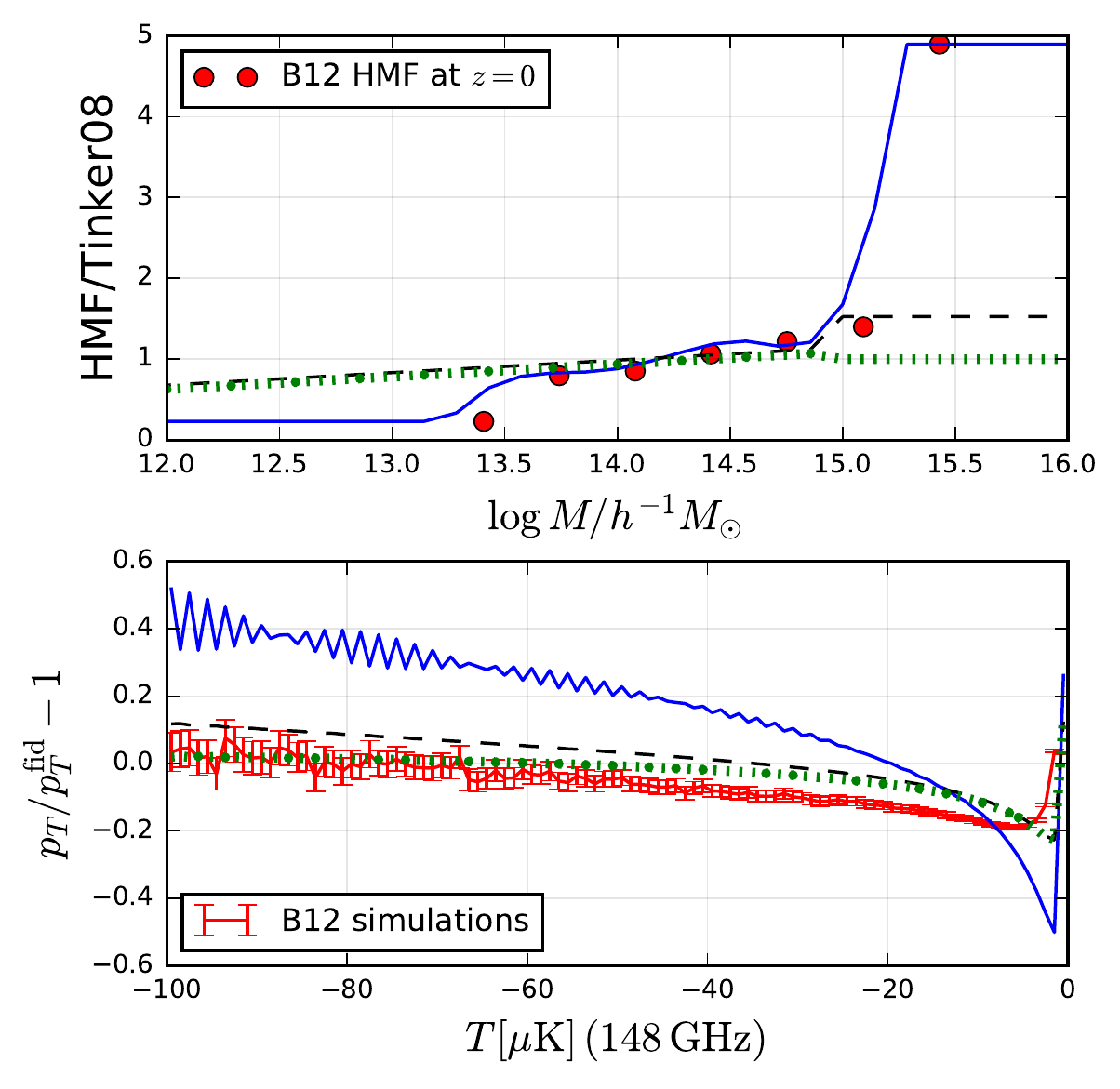}
\caption{\label{fig:hmf_interpolations} Illustration of the impact of the halo mass function
	on the one-point PDF. \emph{Upper panel:} different interpolations of the halo mass function given in~\citetalias{Battaglia2012}.  The red points are the mass function measured at $z=0$ in~\citetalias{Battaglia2012}; the colored curves correspond to various interpolations of these points.  \emph{Lower panel:} corresponding fractional differences to our fiducial result.  The colored curves match the same cases in the top panel. The red points with error bars are identical to those in Fig.~\ref{fig:PDF_sim}.  Modifications of the mass function clearly affect the predicted PDF, and can explain much of the difference between our analytic result and the PDF measured from the hydrodynamical simulations.}
\end{figure}

In Fig.~\ref{fig:hmf_interpolations} we show how different interpolations of the halo mass function (HMF) given in~\citetalias{Battaglia2012} (see their Fig.~11) affect the PDF.  This is compared to the discrepancy found between our fiducial result and the PDF measured from the~\citetalias{Battaglia2012} simulations.  Since only information for redshift $z=0$ is available in Fig.~11 of~\citetalias{Battaglia2012}, we take the ratio between the interpolated halo mass function and the fitting function~\cite{Tinker2008} as constant across redshift.  Although this is likely a gross oversimplification, it can be seen from the figure that reasonable interpolations can already reproduce the observed discrepancy very well.  We thus conclude that our analytic formalism passes this check, although future comparisons with additional hydrodynamical simulations will also be useful.

%%%%%%%%%%%%%%%%%%%%%%%%%%%%%%%%%%%%%%%%%%%%%%
\section{Origin of the Signal}
\label{sec:origin}

\begin{figure}
	\includegraphics[width = 0.5\textwidth]{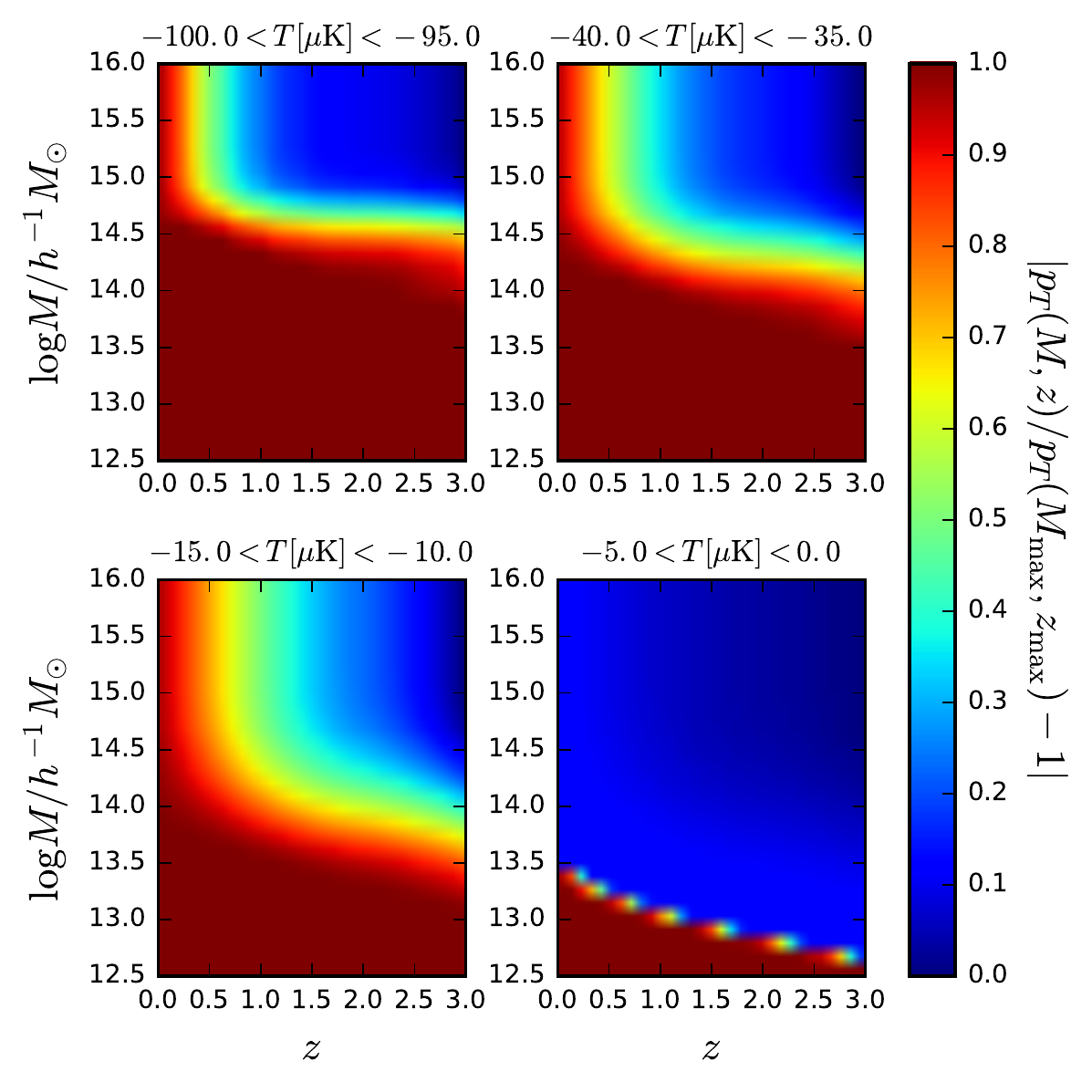}
	\caption{\label{fig:mass_redshift} Mass and redshift contributions. We plot the absolute fractional deviation
		of the PDF in four different temperature bins of width $5\,\mu\rm K$ as a function of the maximum mass and redshift included in the calculation.}
\end{figure}

We now turn to the different contributions to the PDF.
First, we consider cluster mass and redshift.
In Fig.~\ref{fig:mass_redshift} we plot the absolute fractional deviation of the PDF
as a function of the maximum mass and redshift included in the calculation.
These results were obtained using the simplified simulations generated as described in \S\ref{sec:sims}.
It should be noted that care must be taken in interpreting these plots, since overlaps
can shift part of the contribution from a certain $(M,z)$-bin to higher $|T|$ values.
Furthermore, we note that direct comparison with the analogous results given in~\citetalias{Hill2014ACTPDF}
is not possible, because their results included instrumental noise, non-tSZ foregrounds, and beam convolution.

Regarding the mass contributions, we note two general trends in Fig.~\ref{fig:mass_redshift}.
First, as $|T|$ increases, the transition region, i.e., the range of relevant masses
for the specific temperature bin, gets smaller (ignoring the temperature bin that
contains the clear sky fraction for now).
At high $|T|$, the PDF in a given temperature bin is dominated by clusters in a
narrow mass range. On the other hand, for low $|T|$ a variety of sources contributes.
Second, the relevant masses are higher for high $|T|$, which is expected.

Regarding the redshift contributions, the most relevant interval broadens
and shifts to larger redshift as $|T|$ decreases.
For the temperature bin containing the clear sky fraction the behavior is drastically
different, with a very small interval in mass and redshift being the dominant contribution.

\begin{figure}
\includegraphics[width = 0.5\textwidth]{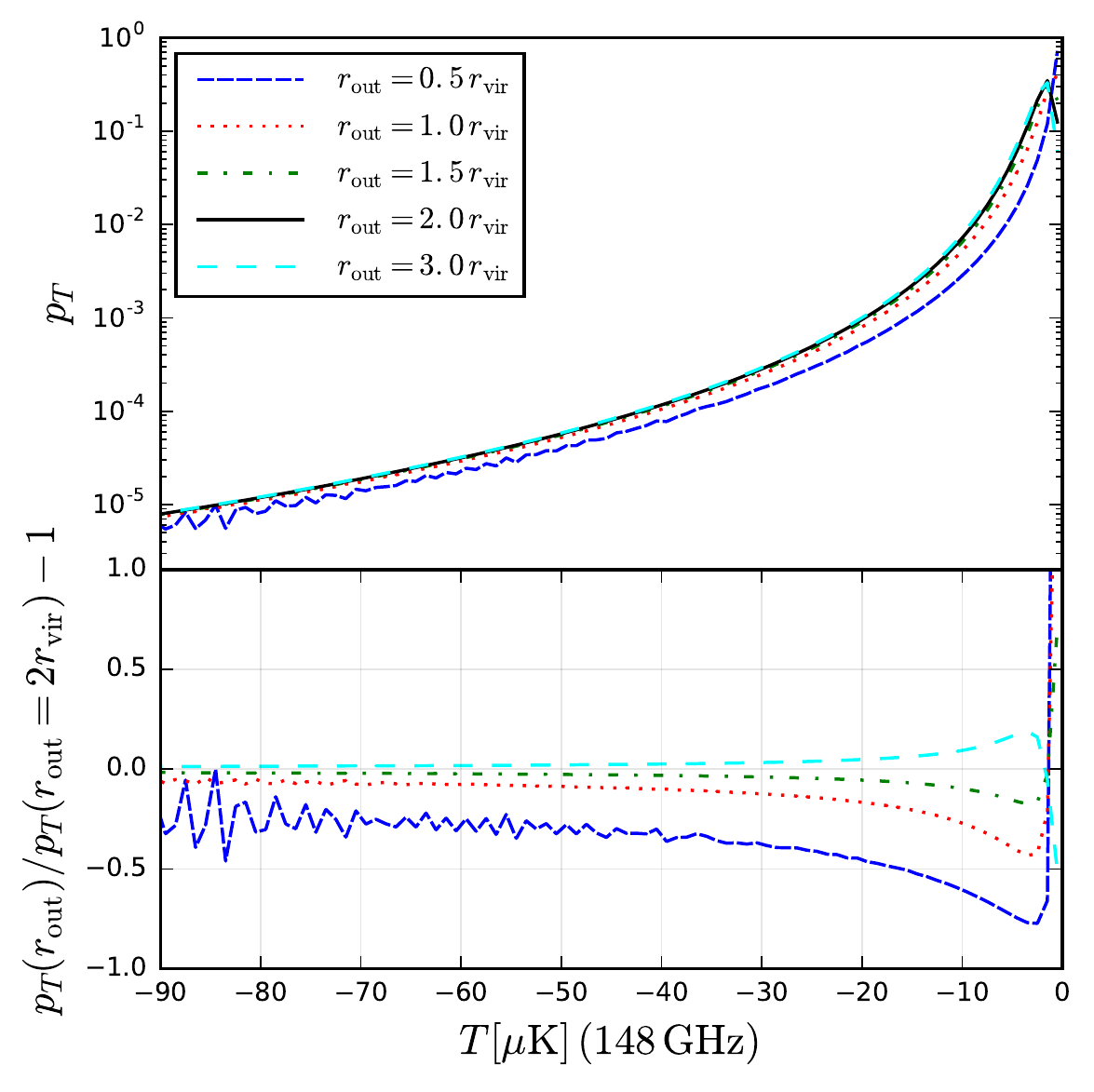}
\caption{\label{fig:radial_cutoffs} Effect of radial cutoff.  The top panel shows the tSZ PDF computed via Eq.~\ref{eq:PDF_clustered} with varying choices of the outer radial cutoff of the pressure profile (as labeled). For the curve corresponding to $r_{\rm out} = 0.5\,r_{\rm vir}$, the ringing
	is due to non-convergence of the FT as mentioned earlier.  The bottom panel shows the fractional difference with respect to our fiducial choice of $r_{\rm out} = 2\,r_{\rm vir}$.}
\end{figure}

Now we consider the effect of the radial cutoff.
In our fiducial computation we considered the $y$-profile up to $r_{\rm out} = 2\,r_{\rm vir}$.
In Fig.~\ref{fig:radial_cutoffs} we show how the choice of this outer radius affects
the PDF.
The effect is largest for the low-$|T|$ regime. These temperature bins receive significant
contributions from the outskirts of clusters. Furthermore, as discussed above, overlaps have
their largest impact on these bins, and reducing the outer radius decreases the amount of overlaps.
The clear sky fraction is increased as $r_{\rm out}$ is decreased, which is intuitively clear.
We note that the choice of $r_{\rm out} = 2\,r_{\rm vir}$ does not correspond to convergence,
since there is still a significant discrepancy to the result obtained with $r_{\rm out} = 3\,r_{\rm vir}$.
However, it is physically not justified to suppose the validity of the pressure profile
fitting function up to infinite radius.  Physically, the virial shock will lead to a sharp decline in the pressure profile at $r \approx 2 - 2.5 r_{200c}$.  
Upon convolution with instrumental noise (as described in \S\ref{sec:exp}) the precise choice of radial
cutoff becomes irrelevant to the PDF prediction, as the extremely small $y$ values in the outskirts are subsumed into the noise.
%\commentleander{we leave consideration of this question to future work}.

%%%%%%%%%%%%%%%%%%%%%%%%%%%%%%%%%%%%%%%%%%%%%%
\section{Parameter Dependence}
\label{sec:parameters}

\begin{figure*}
\includegraphics[width = \textwidth]{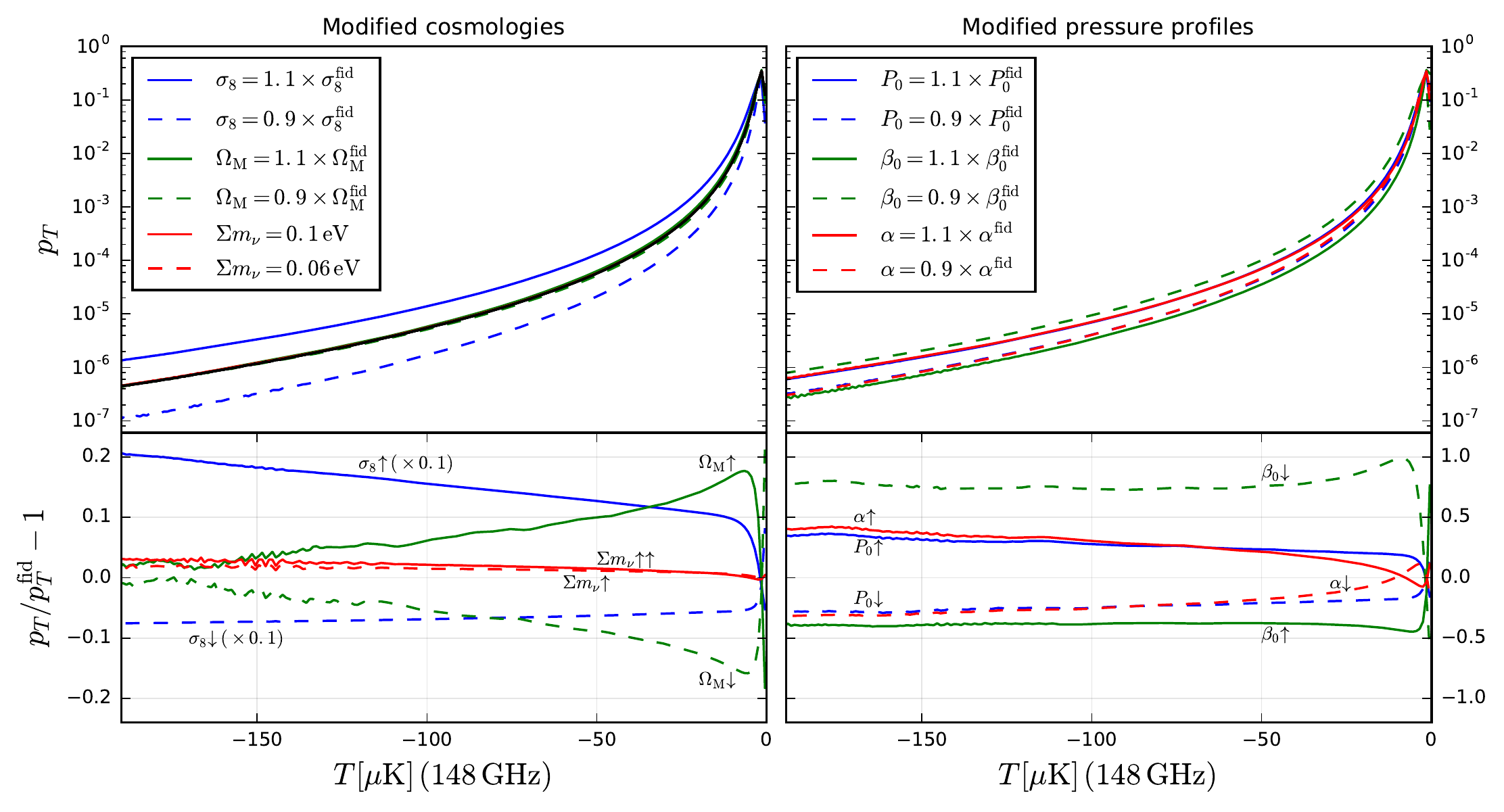}
\caption{\label{fig:cosmology_and_pressure} Effect of varying cosmological (left) and ICM pressure profile (right) parameters. For clarity,
	the residual curves for $\sigma_8$ are reduced by a factor of 10. The wiggles at large $|T|$ are due to the
	angular grid on which the $y$-profiles have been evaluated. Note the different vertical scale in the two
	residual plots.}
\end{figure*}

We now turn to the dependence of the PDF on cosmological and pressure profile parameters.
In Fig.~\ref{fig:cosmology_and_pressure} we plot the effect of changing the cosmological
parameters $\sigma_8$, $\Omega_{\rm M}$, and $\Sigma m_{\nu}$, as well as the pressure profile parameters
$P_0$, $\beta_0$, and $\alpha$ as defined in~\S\ref{sec:theory}.
When considering non-zero neutrino masses we take $N_{\rm eff} = 3.046$.

The impact of changing the cosmology is as follows.  The neutrino mass sum has its largest effect
on high $|T|$ bins, consistent with the fact that it changes the matter power spectrum most on
small spatial scales, which here leads to a suppression in the number of clusters (however, to hold $\sigma_8$ fixed, the initial scalar amplitude $A_s$ must be increased, thus leading to an overall increase in the number of clusters; this is simply an artifact of this choice of cosmological parameters).
The impact of changing $\sigma_8$ is similar, yielding a mild degeneracy with $\Sigma m_{\nu}$.
On the other hand, changing $\Omega_{\rm M}$ has largest impact on low $|T|$ bins.  Thus, degeneracy between $\sigma_8$ and $\Omega_{\rm M}$ could be broken by a measurement of the tSZ PDF over a sufficiently large $|T|$ range.

The pressure profile parameters affect the PDF as follows.
$P_0$ changes the PDF relatively constantly across temperature, which is explained by the fact
that it is an overall normalization of the pressure profile.
Changing the logarithmic slope at large radii, $\beta_0$, produces a similar effect, although with the opposite sign.
Increasing the logarithmic slope at intermediate radii, $\alpha$, decreases the PDF in low $|T|$
bins and increases it in large $|T|$ bins.

%%%%%%%%%%%%%%%%%%%%%%%%%%%%%%%%%%%%%%%%%%%%%%
\section{Including noise and foregrounds}
\label{sec:exp}

Here we consider a representative upcoming CMB experiment and demonstrate the necessity of our improved analytic approach for modeling the tSZ PDF sufficiently accurately to perform unbiased inference from this observable.  In lieu of full parameter forecasts, which require treatment of the covariance matrix and likelihood function, we simply compute the predictions from our full analytic model~(Eq.~\ref{eq:PDF_clustered}) and from the~\citetalias{Hill2014ACTPDF} model~(Eq.~\ref{eq:H14}), and compare these to predicted error bars on a measurement of the tSZ PDF.

Specifically, we consider a measurement with the Simons Observatory (SO), which will cover $f_{\rm sky} = 0.4$ with six frequency channels, reaching a depth and resolution at 145 GHz of $6 \, \mu$K-arcmin and FWHM = 1.4 arcmin, respectively~\cite{SO2018}.\footnote{The sensitivity values considered here are for the ``goal'' configuration presented in Ref.~\cite{SO2018}.}  We include the effects of non-tSZ foregrounds via the post-component-separation Compton-$y$ noise power spectra that are publicly available via the SO Collaboration.\footnote{\url{https://simonsobservatory.org/assets/supplements/20180822_SO_Noise_Public.tgz}}  In particular, we use the $y$ noise power spectrum, $N_{\ell}^{yy}$, derived via the ``standard internal linear combination'', without additional deprojection constraints (see~\cite{SO2018} for further details about the foreground modeling and component separation).  In practice, deprojection options will have to be explored in the foreground cleaning as well, but this is beyond the scope of our work here.

We construct a Wiener filter to optimally weight the harmonic-space tSZ signal via
\begin{equation}
\label{eq.filt}
F_{\ell} = C_{\ell}^{yy} / (C_{\ell}^{yy} + N_{\ell}^{yy}) \,,
\end{equation}
where $C_{\ell}^{yy}$ is the tSZ power spectrum, which we compute using the fiducial model of~\cite{Hill-Pajer2013}.  The filter is smoothly tapered to zero at the boundaries of the multipole range provided in the $N_{\ell}^{yy}$ data file ($\ell_{\rm min} \approx 40$ and $\ell_{\rm max} \approx 8000$).  We apply this filter to the $y$-profiles of all halos in our analytic calculation, which captures the suppression of modes lost due to foregrounds and noise.  Note that the multifrequency information of SO (and Planck, which is also used) allows large-scale tSZ modes to be included (because the CMB can be removed using spectral information), which were lost due to CMB ``noise'' in the single-frequency ACT analysis of~\citetalias{Hill2014ACTPDF}.  Thus the filter extends to lower multipoles than in~\citetalias{Hill2014ACTPDF} (which used a filter originally constructed in~\cite{Wilson2012}).

After calculating the analytic prediction for the filtered tSZ PDF, we convolve the result with a Gaussian noise (+residual foreground) PDF, whose variance is computed via
\begin{equation}
\label{eq.noise}
\sigma_{yy}^2 = \sum_{\ell} \frac{2\ell+1}{4\pi} N_{\ell}^{yy} F_{\ell}^2 p_{\ell}^2 \,,
\end{equation}
where $p_{\ell}$ is the pixel window function (here assumed to be 0.5 arcmin circular pixels, although this has negligible effect).  We then rescale the results from Compton-$y$ to 148 GHz temperature to match those shown elsewhere in the paper (although the application of the filter and noise convolution mean the temperature values are not comparable to those in earlier plots).  We bin the results into bins of width $5\,\mu$K.

\begin{figure}
\includegraphics[width = 0.5\textwidth]{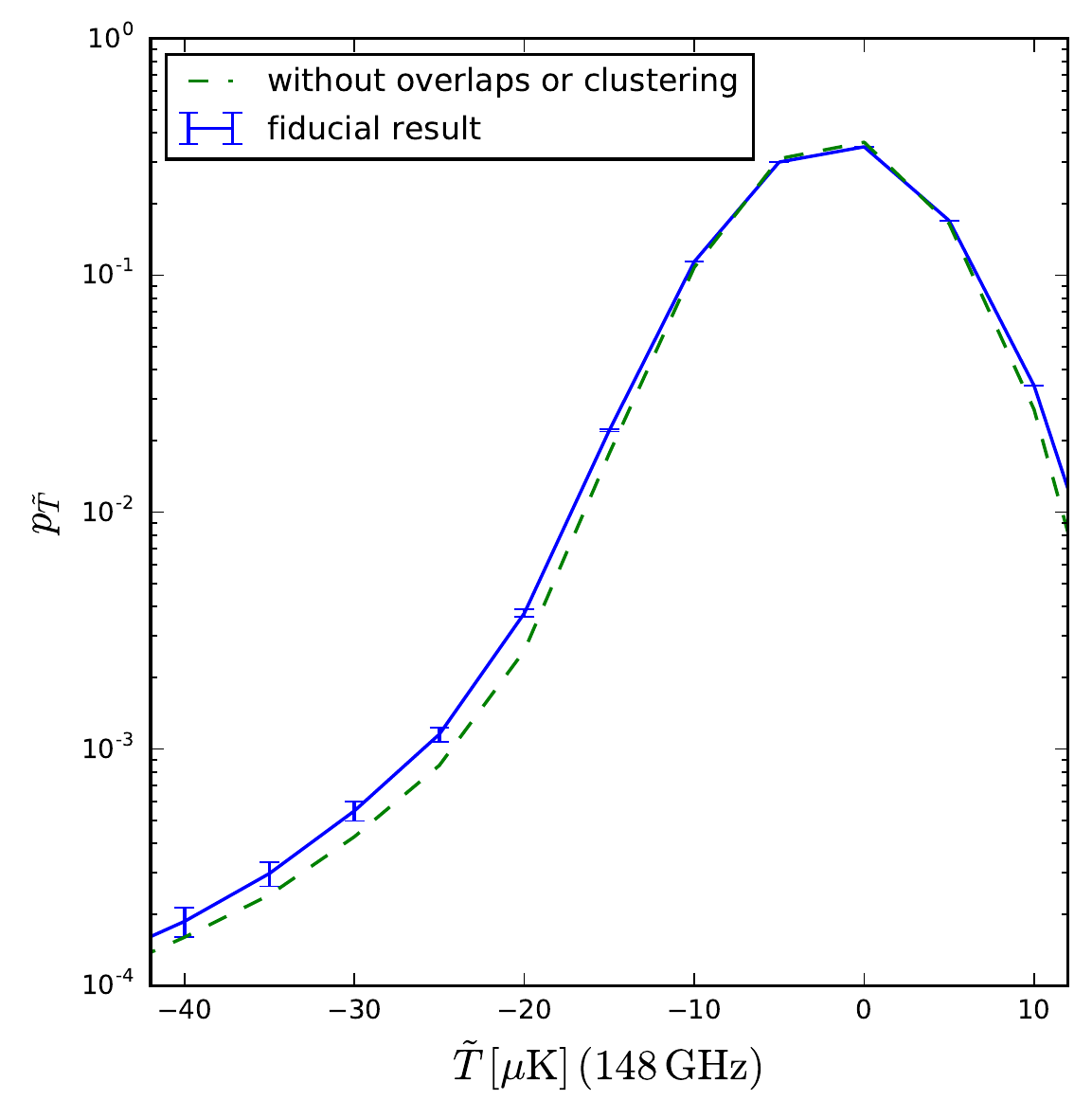}
\caption{\label{fig:with_noise} Difference between our fiducial analytic result (Eq.~\ref{eq:PDF_clustered}; solid blue) and the result neglecting overlaps and halo clustering (i.e., the~\citetalias{Hill2014ACTPDF} model; dashed green), with noise and non-tSZ foregrounds included.  The latter are modeled and propagated through multifrequency component separation via the publicly available Simons Observatory Compton-$y$ noise power spectra.  Note that a Wiener filter has been applied to the $\tilde T$ field here, as denoted by the tilde, and hence the values are not directly comparable to those in other figures.  The convolution with noise and residual foregrounds is responsible for the non-zero PDF values for $\tilde T > 0$.  It is clear that the earlier~\citetalias{Hill2014ACTPDF} model is not sufficiently accurate for SO analysis; the difference between our improved model and the previous model is larger than the error bars in essentially all bins shown.}
\end{figure}

The results are shown in Fig.~\ref{fig:with_noise}.  The solid blue curve shows the prediction from the FT-based formalism developed in this paper (Eq.~\ref{eq:PDF_clustered}).  The dashed green curve shows the prediction from the~\citetalias{Hill2014ACTPDF} model (Eq.~\ref{eq:H14}).  The error bars shown on the blue curve are computed using the diagonal elements of the covariance matrix $C_{ij}$ estimated from
the simplified simulations described in \S\ref{sec:simplified_sims}, but with Poisson distributed cluster numbers, rather than the modified distribution described earlier (which achieved faster convergence at the expense of only capturing the mean correctly), so that the variance is correctly captured.  Specifically, we compute the error on the value of the PDF in the $i$th bin as
\begin{equation}
\sigma_i  = \sqrt{\frac{f_{\rm sky}^{\rm maps}}{f_{\rm sky}^{\rm survey}}} \times \frac{d_{\rm pixel}^{\rm survey}}{d_{\rm pixel}^{\rm maps}} \times \sqrt{C_{ii}},
\end{equation}
where $f_{\rm sky}$ are the sky fractions and $d_{\rm pixel}$ the pixel side lengths.
We take $f_{\rm sky}^{\rm SO} = 0.4$ and $d_{\rm pixel}^{\rm SO} = 0.5\,{\rm arcmin}$;
the simulation parameters are $f_{\rm sky}^{\rm maps} = 6.21\times10^{-3}$ and $d_{\rm pixel}^{\rm maps} = 0.1\,{\rm arcmin}$.
For a survey with the properties described above, the difference between the no-overlaps case
and our fiducial result is considerably larger than the errors for essentially all bins plotted.
We note that the error bars themselves should not be taken at face value,
because significant bin-to-bin correlations exist as discussed in~\citetalias{Hill2014ACTPDF}.
Nevertheless, we conclude that if the earlier model of~\citetalias{Hill2014ACTPDF} were used in an analysis of the tSZ PDF from SO, cosmological and ICM parameter inference would clearly be biased.  With our accurate model in hand, we plan to pursue full parameter forecasts for ongoing and upcoming CMB experiments in future work.

%%%%%%%%%%%%%%%%%%%%%%%%%%%%%%%%%%%%%%%%%%%%%%
\section{Outlook}
\label{sec:outlook}

In this paper we have presented a new analytic model for the tSZ one-point PDF, building upon and substantially improving the model first developed in~\citetalias{Hill2014ACTPDF}.  In particular, by working in Fourier conjugate space, we have shown how to account for effects due to overlaps in the tSZ profiles of halos on the sky, as well as contributions due to the clustering of halos (which arise because of the LOS projection).  For the tSZ PDF, the effects due to overlaps are non-negligible, but the clustering effects are rather small.  We have verified the accuracy of the model via comparison to numerical simulations, both simplified simulations containing randomly distributed clusters and full-scale cosmological hydrodynamics simulations.  However, issues related to the halo mass function in the latter simulations rendered a precise test of the clustering effects challenging; future simulation comparisons will thus be useful.  Finally, we have demonstrated that the use of this more accurate analytic model will be necessary in analyses of the tSZ PDF in upcoming, high-sensitivity CMB data sets.

We anticipate a number of interesting next steps in this line of research.  An obvious first step is to compute the covariance matrix in this formalism and the likelihood function associated with the PDF observable.  Given the challenges observed in this context in~\citetalias{Hill2014ACTPDF}, it may be useful to pursue novel approaches such as likelihood-free inference (although this could render the analytic model redundant)~\cite{Alsing2018}.  We expect that the forecast cosmological constraints using the tSZ PDF will significantly improve upon those for the tSZ power spectrum alone (e.g., as presented in Ref.~\cite{SO2018}).  An optimal combination with constraints from individually detected clusters is clearly also a pressing issue, and will lead to further improvements.

Beyond the tSZ signal, the formalism developed here likely has applications to other cosmological fields.  An obvious candidate is the one-point PDF of the weak lensing convergence field, which has already been investigated in simulations~\cite{Liu2016,Patton2017,Liu-Madhavacheril2018}.  We expect that the clustering effects computed in this paper will be more important for this application than for the tSZ field.  In addition, further development to treat negative-convergence regions (voids) will be necessary.  Nevertheless, a full, non-perturbative model for the one-point PDF of the projected density field is clearly a goal worth pursuing.

%%%%%%%%%%%%%%%%%%%%%%%%%%%%%%%%%%%%%%%%%%%%%%

%%%%%%%%%%%%%%%%%%%%%%%%%%%%%%%%%%%%%%%%%%%%%%
\begin{acknowledgments}
We thank Nick Battaglia and David Spergel for helpful conversations.
JCH is supported by the Friends of the Institute for Advanced Study. 
LT acknowledges support by the Studienstiftung des deutschen Volkes.
KMS was supported by an NSERC Discovery Grant and an Ontario Early Researcher Award.
Research at Perimeter Institute is supported by the Government of Canada
through Industry Canada and by the Province of Ontario through the Ministry of Research \& Innovation.
This is not an official SO Collaboration paper.
\end{acknowledgments}
%%%%%%%%%%%%%%%%%%%%%%%%%%%%%%%%%%%%%%%%%%%%%%

%%%%%%%%%%%%%%%%%%%%%%%%%%%%%%%%%%%%%%%%%%%%%%
\begin{appendix}
%\section{Derivation of Eq.~\ref{eq:PDF_single_slice}}
%\label{app:derivation_PDF_single_slice}
%
%Introduce the auxiliary quantity
%\begin{equation*}
%Y(M,z,\lambda,\theta) \equiv e^{i\lambda y_0(M,z,\theta)} - 1.
%\end{equation*}
%According to Eq.~\ref{eq:def_tildeP},
%\begin{equation}
%\tilde P_{M,z}(\lambda) = \langle \exp i\lambda y(\mathbf{n}) \rangle_{y,{\rm HMF}},
%\end{equation}
%where the average runs over all $y$-signals and samples the halo mass function at the same time.
%We can replace this average by
%\begin{equation*}
%\langle (\cdot) \rangle_{y,{\rm HMF}} = \Omega^{-1}\int_{\Omega} d\mathbf{n}\,\langle(\cdot)\rangle_{\rm HMF}
%\end{equation*}
%Since we are considering an infinitesimal mass-redshift bin, it contains
%at most a single cluster with probability $\Omega dn/d\Omega$, which is located
%at an arbitrary position $\mathbf{n}_{\rm cl}$. We obtain:
%\begin{align}
%\label{eq:PDF_single_slice_expansion}
%\tilde P_{M,z}(\lambda) &= 1 + \left\langle Y(M,z,\lambda,|\mathbf{n}-\mathbf{n}_{\rm cl}|) \right\rangle_{y,{\rm HMF}} \nonumber\\
%						&= 1 + \frac{dn(M,z)}{d\Omega}\int d\mathbf{n}\,Y(M,z,\lambda,|\mathbf{n}-\mathbf{n}_{\rm cl}|) \nonumber\\
%						&= 1 + \frac{dn(M,z)}{d\Omega}\tilde Y(M,z,\lambda).
%\end{align}
%The last line follows from the definition Eq.~\ref{eq:def_tildeY}.
%Since $dn/d\Omega$ is infinitesimal, Eq.~\ref{eq:PDF_single_slice} follows.

\section{Equivalence to Formalism of~\citetalias{Hill2014ACTPDF}}
\label{app:equivalence}

In this appendix we show that our analytic model is equivalent to the formalism
used in~\citetalias{Hill2014ACTPDF} under the assumption that no overlaps occur.
Denote $\int dM\int dz(\chi^2/H)(dn/dM)$ by $\int_{M,z}$ for brevity. The arguments $M$ and $z$ are understood for $y_0(M, z, \theta)$, etc.
Integrating Eq.~(\ref{eq:narrow_bin_expanded}) (which is equivalent to a first order expansion of Eq.~\ref{eq:PDF_unclustered}) over mass and redshift, we obtain
\begin{equation*}
\tilde P(\lambda) = 1 + \int_{M,z}\,\int d\theta\,2\pi\theta(\mathrm{e}^{i\lambda y_0(\theta)}-1).
\end{equation*}
The one-point PDF in $y$-space is given by:
\begin{align}
P(y) &= \delta(y) + \int_{M,z} \int d\theta\,\theta\int d\lambda\,(\mathrm{e}^{i\lambda[y_0(\theta)-y]}-\mathrm{e}^{-i\lambda y}) \nonumber\\
&= \delta(y) + \int_{M,z} \left[ -2\pi\delta(y)\frac{\theta_{\rm max}^2}{2} + 2\pi\frac{\theta_0(y)}{|dy/d\theta_0|}  \right], \nonumber
\end{align}
where we denote the inverse function to $y_0(\theta)$ by $\theta_0(y)$.
The PDF binned into $y_i \leq y \leq y_{i+1}$ is then found as
\begin{align}
\label{eq:old_approach}
p_i &= \int_{y_i}^{y_{i+1}} dy\,P(y) \nonumber\\
&= \delta_i\,\left(1 - \int_{M,z}\pi\theta_{\rm max}^2\right) + \int_{M,z}2\pi\int_{y_i}^{y_{i+1}}dy\,\left\vert\frac{d\theta_0}{dy}\right\vert\theta_0(y) \nonumber\\
&= \delta_i\,(1-F_{\rm clust}) + \int_{M,z} \pi \left[ \theta_0^2(y_{i+1}) - \theta_0^2(y_i)  \right]
\end{align}
where $\delta_i$ equals one if $y = 0$ is contained in the integration interval and zero otherwise.
$\theta_{\rm max} = \theta_{\rm max}(M,z)$ corresponds to the radial cutoff,
so that $1 - F_{\rm clust}$ is the clear-sky fraction.
This is the expression used in~\citetalias{Hill2014ACTPDF}, i.e., Eq.~(\ref{eq:H14}) presented earlier.

\end{appendix}
%%%%%%%%%%%%%%%%%%%%%%%%%%%%%%%%%%%%%%%%%%%%%%

%%%%%%%%%%%%%%%%%%%%%%%%%%%%%%%%%%%%%%%%%%%%%%
%\bibliography{onepoint_paper}

%%%%%%%%%%%%%%%%%%%%%%%%%%%%%%%%%%%%%%%%%%%%%%

\end{document}